\documentclass[runningheads]{llncs}
\usepackage[T1]{fontenc}
\usepackage{graphicx}
\usepackage{lineno}
\usepackage{bm}
\usepackage[hidelinks]{hyperref}
\usepackage{breakurl}
\usepackage{soul}
\usepackage{amsmath,amsfonts}
\usepackage{diagbox}
\usepackage{censor}
\begin{document}
\title{Vestibular schwannoma growth prediction \\from longitudinal MRI \\by time-conditioned neural fields}
\titlerunning{Vestibular schwannoma growth prediction}
%

\author{Yunjie Chen \inst{1} \and
Jelmer M. Wolterink \inst{2} \and
Olaf M. Neve \inst{3} \and
Stephan R. Romeijn \inst{1}  \and
Berit M. Verbist \inst{1}  \and
Erik F. Hensen \inst{3} \and
Qian Tao \inst{4} \and
Marius Staring \inst{1}}
\authorrunning{Y. Chen et al.}
%
\institute{Department of Radiology, Leiden University Medical Center, \\ Leiden, the Netherlands \and
Department of Applied Mathematics, Technical Medical Center, \\ University of Twente, Enschede, the Netherlands \and 
Department of Otorhinolaryngology and Head \& Neck Surgery, \\ Leiden University Medical Center, Leiden, the Netherlands \and
Department of Imaging Physics, Delft University of Technology, \\ Delft, the Netherlands}

\maketitle              

\begin{abstract}
Vestibular schwannomas (VS) are benign tumors that are generally managed by active surveillance with MRI examination. To further assist clinical decision-making and avoid overtreatment, an accurate prediction of tumor growth based on longitudinal imaging is highly desirable. In this paper, we introduce DeepGrowth, a deep learning method that incorporates neural fields and recurrent neural networks for prospective tumor growth prediction. In the proposed method, each tumor is represented as a signed distance function (SDF) conditioned on a low-dimensional latent code. Unlike previous studies that perform tumor shape prediction directly in the image space, we predict the latent codes instead and then reconstruct future shapes from it. To deal with irregular time intervals, we introduce a time-conditioned recurrent module based on a ConvLSTM and a novel temporal encoding strategy, which enables the proposed model to output varying tumor shapes over time. The experiments on an in-house longitudinal VS dataset showed that the proposed model significantly improved the performance ($\ge 1.6\%$ Dice score and $\ge0.20$ mm 95\% Hausdorff distance), in particular for top 20\% tumors that grow or shrink the most ($\ge 4.6\%$ Dice score and $\ge 0.73$ mm 95\% Hausdorff distance). Our code is available at ~\burl{https://github.com/cyjdswx/DeepGrowth}.

\end{abstract}

\section{Introduction}
Vestibular schwannomas (VS) are intracranial tumors arising from the balance and hearing nerves, of which approximately 40\% are progressive and ultimately become life-threatening~\cite{carlson2021vestibular}. In current clinical practice, VS are generally managed by active surveillance with MRI examination and manual tumor diameter measurements~\cite{neve2023automated,kanzaki2003new}. Once significant growth ($>2$mm difference between two consecutive MRI scans) is detected, the tumors are treated with either radiotherapy or surgery~\cite{neve2023automated,marinelli2023size}. However, research shows that although 80\% of VS shows certain growth during observation, only half of them are truly progressive, indicating that many patients suffer from overtreatment~\cite{marinelli2023size}. On the other hand, late treatment of a larger tumor can also damage the prognosis after treatment, which requires a timely clinical decision~\cite{li2015analysis}. Hence, to avoid overtreatment and sequelae associated with the treatment of large tumors, early and precise prediction of tumor growth based on longitudinal imaging is highly desirable. 

Most previous studies on tumor growth prediction leverage recurrent neural networks and generate future tumor shapes directly in the image space. Zhang et al.~\cite{zhang2019spatio} applied a spatio-temporal ConvLSTM for pancreatic tumor growth modeling. Elazab et al.~\cite{elazab2020gp} proposed a 3D GP-GAN that utilizes multiple stacked generative adversarial networks to predict glioma growth. Subsequently, Wang et al.~\cite{wang2022static} applied a Transformer model to longitudinal CT for 4D lung cancer tumor modeling. Although promising results were demonstrated, most models assume unified time intervals between consecutive scans, which is unfortunately uncommon in the clinic. Moreover, future prediction in high-dimensional image space has large memory requirements, which could limit application~\cite{zhang2019spatio}, and may also introduce spatial redundancy that potentially damage performance~\cite{liu2021deep}.

One way to tackle this problem is compressing the input into a low-dimensional latent code utilizing an autoencoder and performing predictions in the latent space~\cite{hu2024complexity,rombach2022high}. In line with this, we propose to perform future tumor prediction with neural field representations~\cite{park2019deepsdf,wiesner2024generative}. The key idea of neural fields is to represent a function describing an image or object in the spatial or spatio-temporal domain as a neural network with trainable weights~\cite{xie2022neural}. The neural network can be conditioned on latent codes to represent a distribution of objects. Recently, Agro et al.~\cite{agro2023implicit} successfully predicted future occupancy maps using spatio-temporal neural fields. However, the method requires sufficient frames over time, while longitudinal medical imaging usually contains only few measurements.

To address these limitations, we propose DeepGrowth, a model that incorporates neural fields and recurrent neural networks for tumor growth prediction. Specifically, DeepGrowth encodes prior images and tumor masks into latent codes and parameterizes the tumor as a signed distance function (SDF). To deal with irregular time intervals between scans, we apply a time-conditioned recurrent module to predict the latent code, on which the reconstruction of the future tumor shape is conditioned. The main contributions of this work are: (1) In contrast to previous studies that perform tumor prediction directly in image space, for the first time, we represent tumor shapes as neural fields and predict the future based on learned latent codes. (2) We introduce a time-conditioned recurrent module with a novel temporal encoding strategy that enables us to query tumor shapes at specific time intervals. (3) The proposed model was evaluated on an in-house longitudinal VS dataset, showing a significantly better performance than other models, in particular for relatively fast growing tumors.

\begin{figure}[tb]
\includegraphics[width=\textwidth]{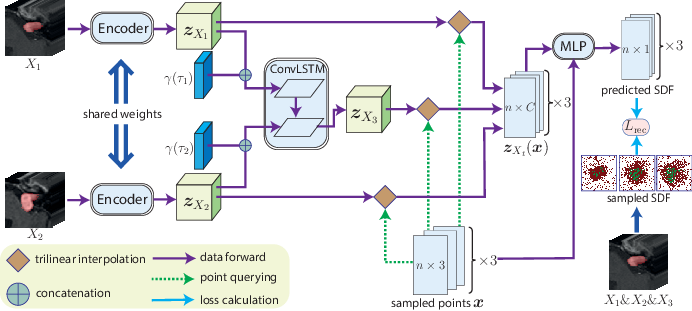}
\caption{The overall architecture of DeepGrowth ($N=3$). Prior scans are encoded into latent codes, which are concatenated with temporal encoding. The MLP reconstructs the future tumor as an SDF conditioned on the output of the ConvLSTM. $L_{\mathrm{rec}}$ is calculated between the predictions and SDF sampled from all three tumor masks.} 
\label{fig:architecture}
\end{figure}

\section{Methods}
Given a patient with $N$ longitudinal images with corresponding segmentations, denoted as $X_t = \{I_t,M_t,D_t\}$, $t=1,2,...,N$, where $I_t$ is the image at time $t$, $M_t$ is the corresponding tumor mask and $D_t$ the normalized scan date ranging from 0 to 1, our goal is to find a function $\Phi$:
\begin{linenomath}
    \begin{align}
    \Phi : \{X_1,X_2,...,X_{N-1},D_N\} \rightarrow M_N.
    \end{align}
\end{linenomath}
Instead of performing prediction directly in image space, we encode $X_t$ into a low-dimensional latent code and predict future by a time-conditioned recurrent module. See Fig.~\ref{fig:architecture} for an overview of the model architecture when $N=3$.

\subsection{3D tumor shape as signed distance function} 
In the proposed model, each tumor is encoded into a low-dimensional latent code, which can be used to condition a neural field for tumor shape reconstruction. More specifically, we concatenate $I_t \in \mathbb{R}^{D \times H \times W}$ and $M_t \in \mathbb{R}^{D \times H \times W}$, and encode them via a convolution-based encoder with a downsampling factor $s$. The latent code is denoted as $\bm{z}_{X_t} \in \mathbb{R}^{C \times d \times h \times w}$, where $d=D/s$, $h=H/s$, $w=W/s$, and $C$ is the feature dimension. Unlike studies that use a single vector to represent the entire object~\cite{park2019deepsdf,wiesner2024generative}, our latent code contains $d \times h \times w$ vectors, encoding the local information in a more expressive representation \cite{peng2020convolutional}.

To reconstruct tumor shapes from the latent code, we represent each tumor shape using an SDF~\cite{park2019deepsdf}. For clarity, we use $c_t$ to denote the tumor contour of $M_t$, which is a closed 2D manifold embedded in 3D space. Hence, for each $M_t$, the SDF of the tumor can be defined as:
\begin{linenomath}
    \begin{align}
        \mathrm{SDF}_{M_t}(\bm{x}) = 
            \left\{
            \begin{array}{ll}
            \min_{u\in c_t}\lVert x-u \rVert_2, & \text{if } x \; \text{inside} \; c_t \\
            0, & \text{if } x \; \text{belonging to} \; c_t \\
            -\min_{u\in c_t}\lVert x-u \rVert_2, & \text{if }  x \; \text{outside} \; c_t 
            \end{array}
            \right.
    \end{align}
\end{linenomath}
where $\bm{x}=(x,y,z) \in \mathbb{R}^3$. Different from voxelized or meshed representations, the SDF and therefore $\bm{x}$ is defined over the entire space. In the proposed model, we approximate the SDF by an MLP $f$. Similar to~\cite{peng2020convolutional,chen2023cones}, we apply a local conditioning strategy, in which $\mathrm{SDF}_{M_t}(\bm{x})$ is conditioned on the local latent code $\bm{z}_{X_t}(\bm{x})$. $\bm{z}_{X_t}(\bm{x})$ is a vector of size $C$ queried from the entire latent code $\bm{z}_{X_t}$ using trilinear interpolation~\cite{peng2020convolutional}. For each point $\bm{x}$, we concatenate the coordinates $\bm{x}$ with $\bm{z}_{X_t}(\bm{x})$ as the input of the MLP, which can then be denoted as:
\begin{equation}
      \mathrm{SDF}_{M_t}(\bm{x}) \approx  f_{\bm{\theta}}(\bm{x},\bm{z}_{X_t}(\bm{x})),
      \label{eq:localSDFmapping}
\end{equation}
where $\bm{\theta}$ are the parameters of the MLP. Hence, each tumor contour is described by the zero-level set of the SDF estimated by the MLP.

\subsection{Time-conditioned recurrent module}
Earlier studies on tumor prediction usually assume unified time intervals between consecutive scans~\cite{wang2022static} and predict a more distant future with additional recurrent steps~\cite{elazab2020gp}. However, patients frequently receive follow-up scans with irregular time intervals. We therefore introduce a time-conditioned recurrent module, which consists of temporal encoding and a small 3D ConvLSTM, to predict future tumor shapes. The 3D ConvLSTM takes the input of $\bm{z}_{X_t}$, $t=1,2,...,N-1$ with the study dates $D_t$, $t=1,2,...,N$ and predicts $\bm{z}_{X_N}$. To better encode the temporal information, we apply sinusoidal functions to the time intervals similar to positional encoding~\cite{mildenhall2021nerf}, which we call temporal encoding. Given the time interval $\tau_i = D_{i+1}-D_i$, where  $i=1,2,...,N-1$, the temporal encoding is expressed as follows:
\begin{equation}
    \gamma(\tau_i) = [\sin(2^{0}\pi \tau_i), \cos(2^{0}\pi\tau_i), \ldots, \sin(2^{l-1}\pi \tau_i), \cos(2^{l-1}\pi\tau_i)],
    \label{eq:positionalencoding}
\end{equation}
where $l$ is the order of the temporal encoding. To avoid overfitting, a dropout layer is added to the temporal encoding. We then concatenate $\gamma(\tau_i)$ to all vectors of $\bm{z}_{X_i}$ as the input of the ConvLSTM. Given the output $\bm{z}_{X_n}$ of the ConvLSTM, we can obtain $\mathrm{SDF}_{M_n}$ of the future tumor via Eq.~(\ref{eq:localSDFmapping}).

\subsection{End-to-end network training}
All components are optimized together end-to-end. For training, we randomly sample $n$ points from each tumor volume as the SDF is defined over the entire space. We apply an $\ell_1$ reconstruction loss that maximizes the similarity between the real SDF and the estimations, as suggested in~\cite{park2019deepsdf}, for all $N$ tumors:
\begin{equation}
L_{\mathrm{rec}} =  \frac{1}{n N} \sum^{N}_{t=1}\sum^{n}_{i=1} \lVert f_{\theta}(\bm{x}_i,\bm{z}_{X_t}(\bm{x}_i)) - \mathrm{SDF}_{M_t}(\bm{x}_i)\rVert_1,
\end{equation}
where $\bm{x}_i$ are the sampled points. To stabilize the training, we apply the $\ell_2$ norm to the latent codes as the regularization:
$L_{\text{reg}} = \frac{1}{N}\sum^{N}_{t=1}\lVert\bm{z}_{X_t}\rVert_2$. As a result, the overall loss function of the proposed model is $L =\lambda_{\text{rec}}L_{\text{rec}} +\lambda_{\text{reg}}L_{\text{reg}}$, where $\lambda_{\text{rec}}$ and $\lambda_{\text{reg}}$ are the weights of each loss function.

\begin{table}[tb]
\centering
\caption{Quantitative comparison results on a vestibular schwannoma dataset using 5-fold cross-validation. The mean and standard deviation of Dice, 95\% HD, and RVD are reported. The highest values per column are indicated in bold; $\dagger$ indicates a significant difference (\(p < .05\)) compared to the proposed method.}
\begin{tabular}{l|c|c|c|c}
\hline
Method &\#params & Dice $\uparrow$ & 95\% HD (mm) $\downarrow$ & RVD $\downarrow$ \\
\hline
Stable tumor &  & $0.766 \pm 0.143^{\dagger} $& $1.95 \pm 2.55^{\dagger}$ & $\bm{0.490 \pm 2.99}$\\
\hline
ST-ConvLSTM~\cite{zhang2019spatio} & 0.6 M & $0.758 \pm 0.141^{\dagger}$& $2.07 \pm 2.65^{\dagger}$ & $0.611 \pm 3.62^{\dagger}$\\
\hline
3D ConvLSTM~\cite{shi2015convolutional} & 4.4 M & $0.784 \pm 0.139^{\dagger}$ & $1.91 \pm 2.50^{\dagger}$ & $0.564 \pm 3.47^{\dagger}$\\
\hline
DeepGrowth (proposed) & 4.9 M & $\bm{0.800 \pm 0.115}$ & $\bm{1.71 \pm 2.23}$ & $0.521 \pm 3.48$\\
\hline
\end{tabular}
\label{tb:quantitative_results}
\end{table}

\begin{table}[tb]
\centering
\caption{Quantitative comparison results of the top 20\% fastest growing or shrinking VS using 5-fold cross-validation. The mean and standard deviation of Dice, 95\% HD and RVD are reported. The highest values per column are indicated in bold; $\dagger$ indicates a significant difference (\(p < .05\)) compared to the proposed method.}
\begin{tabular}{l|c|c|c|c}
\hline
Method &\#params & Dice $\uparrow$ & 95\% HD (mm) $\downarrow$ & RVD $\downarrow$\\
\hline
Stable tumor &  & $0.697 \pm 0.182^{\dagger} $& $4.18 \pm 3.30^{\dagger}$ & $0.413 \pm 0.323^{\dagger}$\\
\hline
ST-ConvLSTM~\cite{zhang2019spatio} & 0.6 M & $0.707 \pm 0.188^{\dagger}$& $4.28 \pm 3.42^{\dagger}$ & $0.398\pm 0.470$\\
\hline
3D ConvLSTM~\cite{shi2015convolutional} & 4.4 M & $0.736 \pm 0.176^{\dagger}$ & $3.87\pm 3.22^{\dagger}$ & $0.366 \pm 0.332$\\
\hline
DeepGrowth (proposed) & 4.9 M & $\bm{0.782 \pm 0.120}$ & $\bm{3.14 \pm 2.22}$ & $\bm{0.321 \pm 0.315}$\\
\hline
\end{tabular}
\label{tb:quantitative_results_top20p}
\end{table}

\section{Experiments}
\subsection{Dataset}
To evaluate the proposed method, 131 vestibular schwannoma patients were selected from our previous study~\cite{neve2022fully,neve2023automated}. Each patient in the dataset has three consecutive contrast enhanced T1 (T1ce) scans, separated by 87 to 2157 days. The spatial resolution of the T1ce ranges from $0.254 \times 0.254 \times 0.81$ mm to $1.17 \times 1.17 \times 1.20$ mm and the in-plane resolution ranges from $256 \times 192$ to $640 \times 520$. Of all scans the tumor masks were generated using a segmentation model developed in our previous study based on nnUNet~\cite{isensee2021nnu,neve2022fully}. We aligned all scans of each patient by rigid registration using elastix~\cite{klein2009elastix}. All images were then resampled to an isotropic resolution of $0.58 \times 0.58 \times 0.58$ mm. To avoid the influence of background, $64 \times 64 \times 64$ cropping was performed around the centroid of the tumor. The intensities of T1ce were normalized to $[-1,1]$.

\subsection{Implementation details}
We adapted a 3D U-Net from~\cite{peng2020convolutional} as the encoder with two extra convolutional blocks for downsampling. The ConvLSTM in the time-conditioned recurrent module consists of three 32-channel layers and the MLP contains five 64-channel layers with sine as the activation function~\cite{sitzmann2020implicit}. Due to the limited dataset size and diversity in tumor growth trends, we perform five-fold cross-validation and report the average results of the five folds to avoid bias. We set the downsampling factor $s=4$ and temporal encoding order $l=6$ for best performance (see Section~\ref{sec:ablation}). Little difference was observed between different loss weights, which were set to $\lambda_{\text{rec}} = 1.0$ and $\lambda_{\text{reg}} = 0.1$. The model was optimized using Adam with an initial learning rate of $1e-4$. During inference, the tumor masks were generated from the zero level-set of the predicted SDF and evaluated using the Dice, 95\% Hausdorff distance (95\% HD), and relative volume difference (RVD). All experiments were conducted using Python 3.10 and PyTorch 1.12.1 on a machine equipped with Nvidia Quadro RTX 6000 and Nvidia Tesla V100 GPUs.

\begin{figure}[tb]
\includegraphics[width=\textwidth]{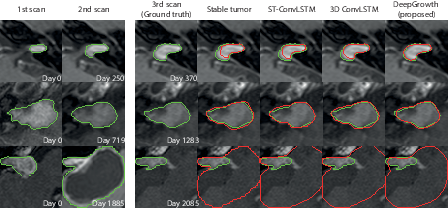}
\caption{Example results of the different models. The first two columns are the input of the models, followed by the ground truth in the third column, and model predictions in subsequent ones. Predicted tumors are depicted in red and the ground truths in green. The dates are the study dates. The last row depicts a tumor that suddenly shrank after the second scan, which was difficult to predict for all models.} 
\label{fig:examples}
\end{figure}

\subsection{Future tumor shape prediction}
We first evaluate the proposed model by predicting the third future tumor shape from the first two scans and time intervals. We compare our model against three baselines. The first baseline assumes the tumor remains stable after the second scan, which is reasonable due to the slow growth of VS, so we simply take the tumor mask from the second time point as a prediction, which we call "stable tumor" in the experiments. The second and third baselines are two ConvLSTM-based models: ST-ConvLSTM~\cite{zhang2019spatio} and 3D ConvLSTM~\cite{shi2015convolutional}. ST-ConVLSTM is a smaller 2D model where we use the same architecture as described in the original paper. 3D ConvLSTM, which contains a comparable number of parameters to the proposed model, consists of three layers with (64, 128, 64) channels respectively. Unlike the original papers that use the $\ell_1$ loss to train the model to generate binary maps, we used a weighted sum of Dice loss and binary cross-entropy loss, which performed better on our data, to train the baselines. Wilxocon signed rank tests were performed between the proposed model and each baseline. 

The quantitative results are listed in Table~\ref{tb:quantitative_results} with visualizations in Fig.~\ref{fig:examples}. The proposed model performed significantly better than all baselines in terms of Dice and 95\%HD. The proposed model obtained a higher RVD due to an extreme outlier (see last row in Fig.~\ref{fig:examples}). When removing this outlier, the proposed method obtained an RVD of $0.218 \pm 0.248$, which outperformed all baselines ($0.229 \pm 0.230$, $0.296 \pm 0.304$ and $0.261 \pm 0.266$, respectively). 

We noticed that the stable tumor method obtained comparable quantitative scores, which is on par with the fact that many VS grow slowly or even remain stable. Focusing on the top 20\% of tumors that grow (or shrink) the most, see Table~\ref{tb:quantitative_results_top20p}, we observe a larger gap between the proposed model and the baselines, indicating the improved capability of modeling tumor growth.

\begin{table}[tb]
\centering

\caption{Quantitative results of top 20\% growers when varying temporal encoding.}
\begin{tabular}{l|c|c|c|c}
\hline
methods &  order & Dice $\uparrow$ & 95\% HD (mm) $\downarrow$ & RVD $\downarrow$\\
\hline
w/o time & & $0.765\pm 0.143 $& $3.37 \pm2.49$ & $0.341 \pm 0.314$\\
with time   & & $0.774 \pm 0.126$ & $3.23 \pm 2.50$ & $0.316 \pm 0.278$\\
\hline
 & $l=4$&  $0.773 \pm 0.144$ & $3.33 \pm 2.53$ & $0.316 \pm 0.306$\\
 time + temporal encoding& $l=6$&  $\bm{0.782 \pm 0.119}$ & $\bm{3.14 \pm 2.22}$ & $
0.321 \pm 0.315$\\
 & $l=8$& $0.773 \pm 0.142$ & $3.23 \pm 2.39$ & $\bm{0.315 \pm 0.363}$\\
\hline
\end{tabular}
\label{tb:timeinfo_ablation}
\end{table}

\begin{figure}[tb]
\includegraphics[width=\textwidth]{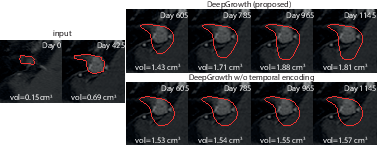}
\caption{Querying the proposed model at different time points (increments of 180 days). We overlaid predictions on $I_2$ for visualization in columns 3-6. The proposed model can output varied tumor shapes given different time intervals, while the model without temporal encoding outputs almost the same results regardless of the time intervals.} 
\label{fig:examples_of_later_tumor}
\end{figure}

\subsection{Ablation study}
\label{sec:ablation}
To examine the impact of temporal encoding, we trained two additional models: one without time factors at all, and one using time intervals $\tau_i$ directly as suggested in~\cite{zhang2019spatio}. We also compare the models using different orders $l$ for temporal encoding. The results of the top 20\% growers are shown in Table~\ref{tb:timeinfo_ablation}. Direct use of $\tau_i$ barely improved results, while temporal encoding improved the results for all metrics. Best results were obtained for $l=6$, with higher $l$ leading to overfitting. 

As our model allows us to query arbitrary future time points, we show predictions given different $\tau_2$ (with a step of 180 days) in Fig.~\ref{fig:examples_of_later_tumor}. We can see that the model using $\tau_i$ without temporal encoding outputs almost the same results regardless of the time intervals. On the contrary, by using temporal encoding, the proposed model can output varied tumor shapes given different time intervals, from which we can view how tumors grow over time.

Models using high-resolution feature maps were more difficult to train, while lower-resolution feature maps potentially degraded performance due to lowered expressive capability~\cite{hu2024complexity,rombach2022high}. We, therefore, varied the downsampling factors $s$, see Table~\ref{tb:downsamplingablation}, and concluded that $s=4$ resulted in the best performance.

\begin{table}[tb]
\centering
\caption{Quantitative results of DeepGrowth using different downsampling factors.}
\begin{tabular}{c|c|c|c}
\hline
downsampling factor s &  Dice $\uparrow$ & 95\% HD (mm) $\downarrow$& RVD $\downarrow$\\
\hline
s=1 & $0.788 \pm 0.127$ & $1.87 \pm 2.39$ & $0.544 \pm 3.68$\\
s=2 & $0.796 \pm 0.122$ & $1.78 \pm 2.40$ & $0.577 \pm 4.18$\\
s=4 & $\bm{0.800 \pm 0.115}$ & $\bm{1.71 \pm 2.23}$ & $
\bm{0.52 \pm 3.48}$\\
s=8 & $0.784 \pm 0.125$ & $1.85 \pm 2.35$ & $0.598\pm 4.10$\\
\hline
\end{tabular}
\label{tb:downsamplingablation}
\end{table}

\section{Discussion and Conclusion}
In this paper, we proposed DeepGrowth, a deep learning model that incorporates neural fields and recurrent neural networks for tumor growth prediction. Unlike conventional models that predict image or segmentation masks directly in the image space~\cite{zhang2019spatio,elazab2020gp}, we encode tumors into a latent space and predict future latent codes. The future tumor shape is reconstructed as the zero-level set of an SDF conditioned on the predicted latent code via an MLP. A comparison on a longitudinal VS dataset showed improved performance of the proposed model, in particular for more challenging growing or shrinking tumors. We applied temporal encoding to the study intervals, which helped the model to encode time information and output varied tumor shapes given different time intervals. However, it remains to be investigated if tumor growth derived from our predictions can be used to aid clinical decision making. In conclusion, we showed that neural fields hold great promise for information compression, which can facilitate longitudinal tumor modeling.

\begin{credits}
\subsubsection{\ackname} This study was supported by the China Scholarship Council (grant 202008130140), and by an unrestricted grant of Stichting Hanarth Fonds, The Netherlands (project MLSCHWAN).

\subsubsection{\discintname}
The authors have no competing interests to declare that are relevant to the content of this article. 
\end{credits}
%
%
%
%
\bibliographystyle{splncs04}
\bibliography{references}

\end{document}